\documentclass[]{aastex631}

\graphicspath{{./}{figures/}}
\usepackage{hyperref}
\usepackage{float}

\begin{document}

\title{Ultra-high Energy $\gamma$ rays from Past Explosions in our Galaxy}

\author[0000-0002-0395-4868]{Alokananda kar}

\email{alokanandakar@gmail.com}
\author[0000-0002-1188-7503]{Nayantara Gupta}
\affiliation{Raman Research Institute \\
C. V. Raman Avenue, 5th Cross Road, Sadashivanagar, Bengaluru, Karnataka 560080, India}
\email{nayan@rri.res.in}

\begin{abstract}
The discovery of the sources of ultra-high energy photons by High-Altitude Water Cerenkov Gamma ray Observatory and Large High Altitude Air Shower Observatory in our Galaxy has revolutionised the field of gamma ray astronomy in the last few years. These emissions are sometimes found in the vicinity of powerful pulsars or supernova remnants associated with giant molecular clouds. Inverse Compton emission by shock accelerated electrons emitted by pulsars and proton-proton interactions of shock accelerated protons emitted by supernova remnants with cold protons in  molecular clouds are often identified as the causes of these emissions. In this paper we have selected two ultra-high energy photon sources LHAASO J2108+5157 and LHAASO J0341+5258 which are associated with giant molecular clouds, but no powerful pulsar or supernova remnant has been detected in their vicinity. We have proposed a scenario where shock accelerated electrons and protons are injected in the local environment of these sources from past explosions, which happened thousands of years ago. We show that the observed ultra-high energy photon flux can be explained with the secondary gamma rays produced by the time evolved relativistic electron and proton spectra.
 
\end{abstract}

\keywords{High-energy astrophysics(739) -- Gamma rays (637) -- Gamma ray sources (633)}

\section{Introduction} \label{sec:intro}

The field of gamma ray astronomy has been rapidly progressing in the past ten years due to the successful operation of many ground based detectors
 like \href{https://www.mpi-hd.mpg.de/hfm/HESS/}{H.E.S.S} (High Energy Stereoscopic System), \href{https://magic.mpp.mpg.de/}{MAGIC} (Major Atmospheric Gamma Imaging Cherenkov) telescope, \href{https://www.icrr.u-tokyo.ac.jp/em/}{Tibet}, \href{https://www.hawc-observatory.org/}{HAWC} (High-Altitude Water Cherenkov) Gamma ray Observatory,  \href{http://english.ihep.cas.cn/lhaaso/}{ LHAASO} (Large High Altitude Air Shower Observatory) and space based detectors   \href{http://www-glast.stanford.edu/}{Fermi LAT}, \href{http://agile.rm.iasf.cnr.it/}{AGILE}.
With the detection of gamma rays of energy several hundreds of TeV to PeV energy in our Galaxy, ultra-high energy (UHE) gamma ray astronomy has emerged as an active area of research. 
These gamma rays are attenuated by cosmic microwave background (CMB) and infrared background (IRB), due to this reason it is challenging to detect UHE gamma ray sources outside our Galaxy.
Gamma rays are messengers of cosmic ray acceleration sites. Thus by detecting the highest energy gamma rays it would be possible to reveal the highest energy cosmic ray acceleration sites within our Galaxy.
\par
The first catalog of gamma ray sources above 56 TeV and 100 TeV detected by HAWC \citep{PhysRevLett.124.021102} shows nine sources above 56 TeV which have at least one pulsar within $0.5^{\circ}$ of HAWC high energy location. 
 Their spatial extensions are much larger than pulsar wind nebulae (PWNe) detected in X-rays, which indicates there could be TeV halos. The gamma rays could be produced in inverse Compton (IC) emission by relativistic electrons escaping from PWNe which are powered by pulsars. This scenario has been used to explain even 100 TeV gamma ray emission detected by HAWC \citep{breuhaus2021ultra}. In this case the IC emission happens in the Klein-Nishina regime and the gamma ray spectrum is softer compared to that in the Thomson regime.
 
 The three gamma ray sources above 100 TeV listed in the first HAWC catalog may also result from proton-proton interactions, as cosmic ray protons of PeV energy escaping from supernova remnants (SNRs) can efficiently produce gamma rays of 100 TeV energy by interacting with cold protons in giant molecular clouds (GMCs). 
\par
The quest for finding the acceleration sites of cosmic rays and the highest energy of the Galactic cosmic rays has led to the discovery of a dozen of UHE gamma ray sources in our Galaxy \citep{cao2021ultrahigh} by LHAASO. Many of these sources are associated with pulsars or supernova remnants. The curvature in the gamma ray spectrum at the highest energy could be an indication of attenuation by background photons. The curvature in gamma ray spectrum may also result from exponential cut-off in parent proton or electron spectrum.
Below we discuss about two UHE gamma rays sources detected by LHAASO. They have been modelled earlier by the LHAASO Collaboration using leptonic and hadronic emission in a steady state model. Since, no PWN or SNR has been detected in their vicinity, it is hard to explain the origin of the constant supply of parent cosmic rays which produce the observed UHE gamma rays. We propose that shock accelerated electrons and protons were injected from explosions more than several thousands of years ago at the locations of the UHE gamma ray sources. The highly relativistic electrons have been losing energy by synchrotron, bremsstrahlung and IC emission and the protons have been losing energy in proton-proton interactions inside GMCs. The time evolved secondary gamma ray spectrum at present day can explain the UHE gamma ray data recorded by LHAASO.

\subsection{LHAASO J2108+5157} \label{sec:source1}

LHAASO $J2108+5157$ is an UHE gamma ray source, discovered by \cite{2021} after analysing the data from LHAASO-KM2A for 308.33 live days. This point-like source is located at  R.A.=$312.17   \pm 0.07_{stat}$ , $Dec. = 51.95 \pm 0.05_{stat}$. This source is located near the center of a GMC $[MML2017]4607$( \citet{miville2017physical}), which is within the upper limit of the extension of LHAASO J2108+5157. The average angular radius
of the GMC [MML2017]4607 is 0.236$^\circ$ with a mass of 8469 \(M_\odot\) at a distance of 3.28 kpc. The number density of particles in this GMC is estimated to be 30 cm$^{-3}$. Due to this spatial correlation a hadronic origin of UHE gamma rays has been speculated, however leptonic origin cannot be ruled out.
The absence of any X-ray counterpart within 0.26$^\circ$ and any very high energy gamma ray counterpart within a radius of 0.5$^\circ$ of the source has been reported. Though there is a high energy gamma ray source 4FGL J2108.0+5155e spatially coincident with LHAASO J2108+5157, its flux is 10 times lower than the UHE gamma ray source. Based on the catalog of SIMBAD2, \cite{2021} searched for possible cosmic accelerators within 0.8$^\circ$ from the center of the LHAASO source. Though no SNR or PWN has been found, two young star clusters Kronberger 82 \citep{kronberger2006new} and Kronberger 80 \citep{kharchenko2016global} have been located. These star clusters could be cosmic ray acceleration sites.

\subsection{LHAASO J0341+5258}
LHAASO J0341+5258 is an extended $\gamma$ ray source discovered by  \citet{cao2021discovery} after analysing the data recorded by LHAASO KM2A over 308.33 days. It is located in the galactic plane, with best fit position at R.A.= $55.34 \pm 0.11$ and Dec= $52.97 \pm 0.07$. The CO-line survey done by Milky  Way Imaging Scroll Painting project \citep{su2019milky} shows that molecular gas partially overlaps with LHAASO J0341+5258. The total mass of gas within 1$^\circ$ of the source is $10^3 M_{\sun}$ if its distance is considered as 1 kpc. Assuming the average cloud thickness is 0.5$^\circ$, the density of particles is estimated as 50 cm$^{-3}$ in the cloud. There are four X-Ray sources within 0.6$^\circ$ of this LHAASO source, namely 2RXS J034125.8+525530, 2RXS J033928.5+530720, 2RXS J034316.5+524331 and 2RXS J034203.0+532329. The positions of 2RXS J033928.5+530720 and 2RXS J034316.5+524331 are coincident with 2SXPS 172133 and 2SXPS 171354 in the 2RXPS Swift X-ray telescope point source catalogue \citep{evans2019vizier}. \citet{cao2021discovery} found that the archived Chandra ACIS observation with ID 16828 partially overlaps with LHAASO J0341+5258. Fermi LAT detected 4FGL J0340.4+5302 is the nearest GeV gamma ray source located within the extension of LHAASO J0341+5258.
Although, the pulsar halo model has been discussed by \citet{cao2021discovery}, due to the absence of any powerful young pulsar in the vicinity of LHAASO J0341+5258 the leptonic scenario of UHE gamma ray production remains questionable. Also, no SNR has been located in the vicinity of LHAASO J0341+5258.

\section{Modeling of Spectral Energy Distribution (SED) of Photons}
We have assumed there were explosions thousands of years ago, which lasted for a year, at the locations of LHAASO J2108+5157 and LHAASO J0341+5258, which injected shock accelerated electrons and protons. We have used a lepto-hadronic model which includes both leptonic and hadronic energy losses to calculate the spectral energy distributions of photons.
The GAMERA code \citep{Hahn:2016CO} has been used, which solves the transport equation to generate the time evolved electron and proton spectra. It also calculates the resulting radiation spectra.
The transport equation for electrons or protons is given by
\begin{equation} \label{trans}
\frac{\partial{N}}{\partial{t}}=Q(E,t)-\frac{\partial{(bN)}}{\partial{E}}-\frac{N(E,t)}{t_{diff}}
\label{eq:trnsp}
\end{equation}
 where $Q(E,t)$ is the injection spectra and $b=b(E,t)$ represents the energy loss of injected particles. $t_{diff}$ denotes the time-scale on which the particles (electrons and protons) escape from the region of explosion due to diffusion. $N(E,t)$ is the resultant particle spectra at any time t.

 We have assumed a power law injection spectra of electrons and protons. For leptonic modeling we have included synchrotron, bremsstrahlung, IC scattering of photons from CMB and radiation fields in Galaxy in the transport equation given in Eq.(\ref{eq:trnsp}) by the term $b(E,t)$. In case of protons, the energy loss is accounted for by proton-proton interactions.
GAMERA uses the full Klein Nishina cross section for IC scattering from \cite{blumenthal1970bremsstrahlung} to calculate the IC photon flux radiated by the relativistic electrons. Electrons also produce bremsstrahlung radiation after interacting with molecular clouds. The code calculates this radiation for both electron-electron and electron-ion interactions \citep{baring1999radio}. For proton-proton interactions GAMERA uses the parametrisation developed by \cite{kafexhiu2014parametrization}. The hadronic interaction model GEANT 4.10.0 has been used in our calculations. \par
The diffusion loss of particles has been included through the diffusion time-scale $t_{diff}$ in the third term of the transport equation. For cosmic ray propagation in our Galaxy, $t_{diff}$ can be expressed as,
\begin{equation}\label{diff}
    t_{diff}=\frac{H^2}{D}.
\label{eq:difftime}    
\end{equation}
\begin{figure*}
\gridline{\fig{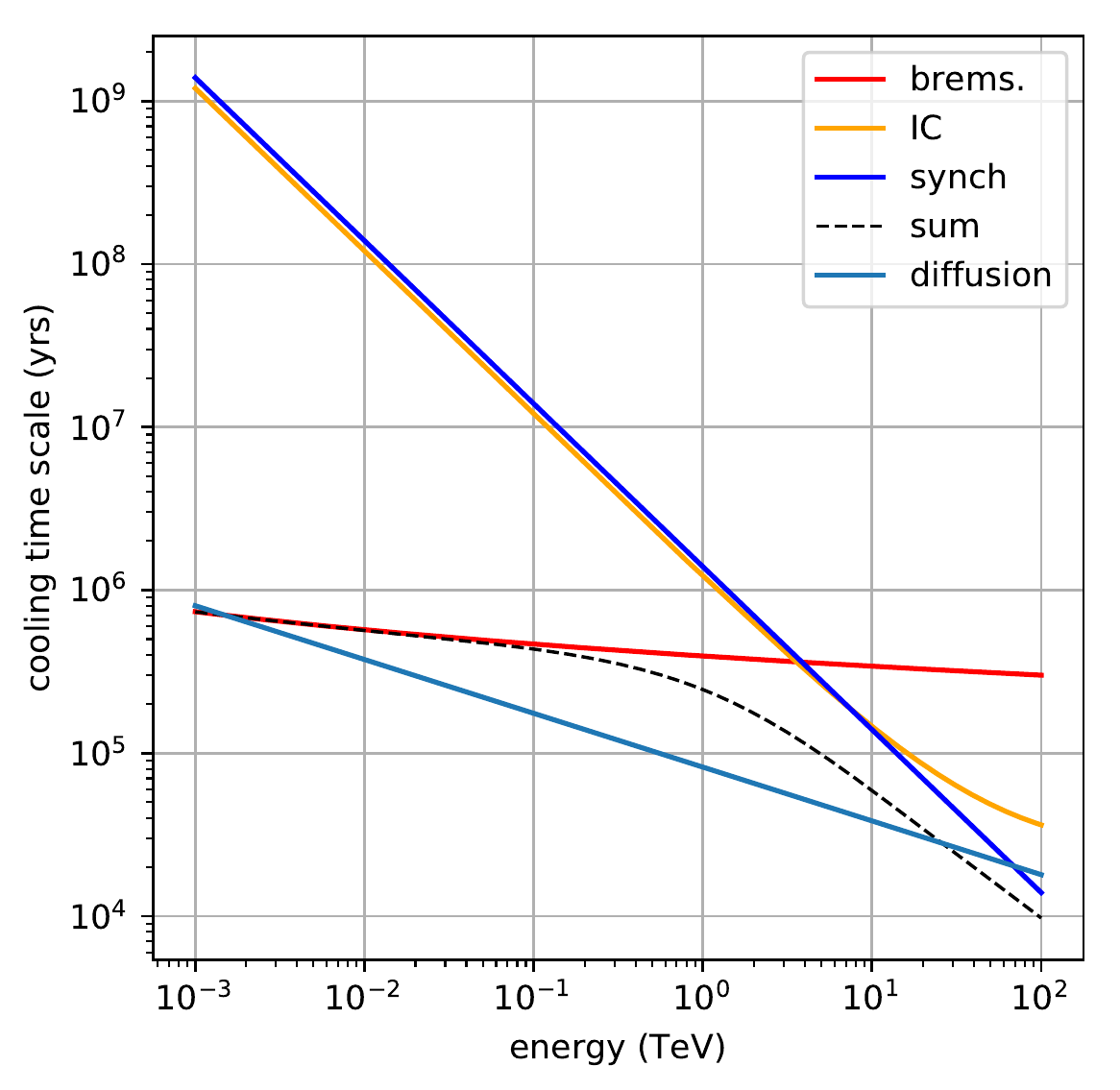}{0.5\textwidth}{(a)}
          \fig{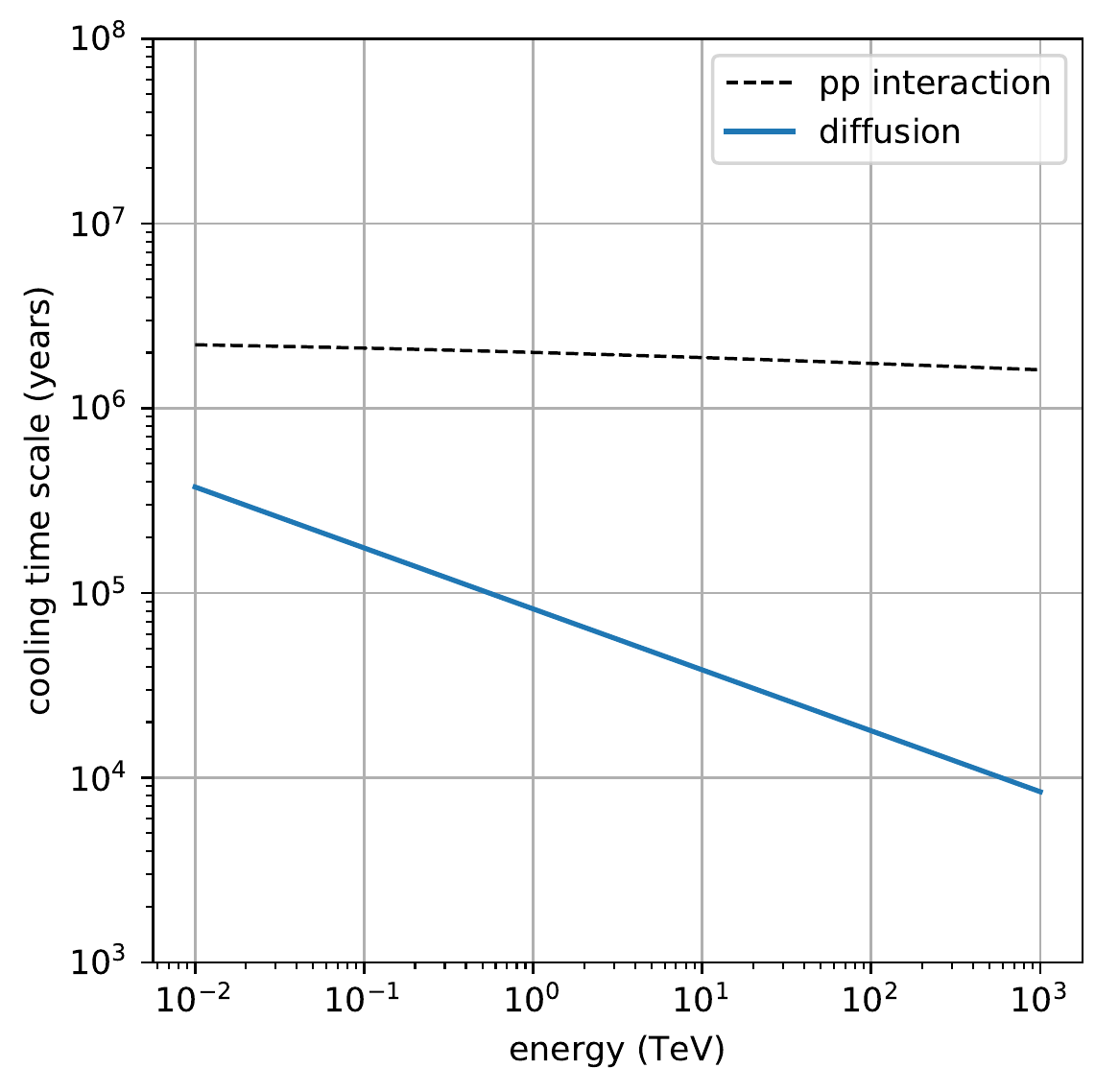}{0.5\textwidth}{(b)}
}
\caption{ Diffusion and cooling time-scales of (a) electrons due to synchrotron, inverse-Compton (IC) and bremsstrahlung emission, `sum' denotes the total cooling time-scale after including all these processes (b) protons due to proton-proton interactions for LHAASO J2108+5157
\label{fig:cooling1}}
\end{figure*}
\begin{figure*}
\gridline{\fig{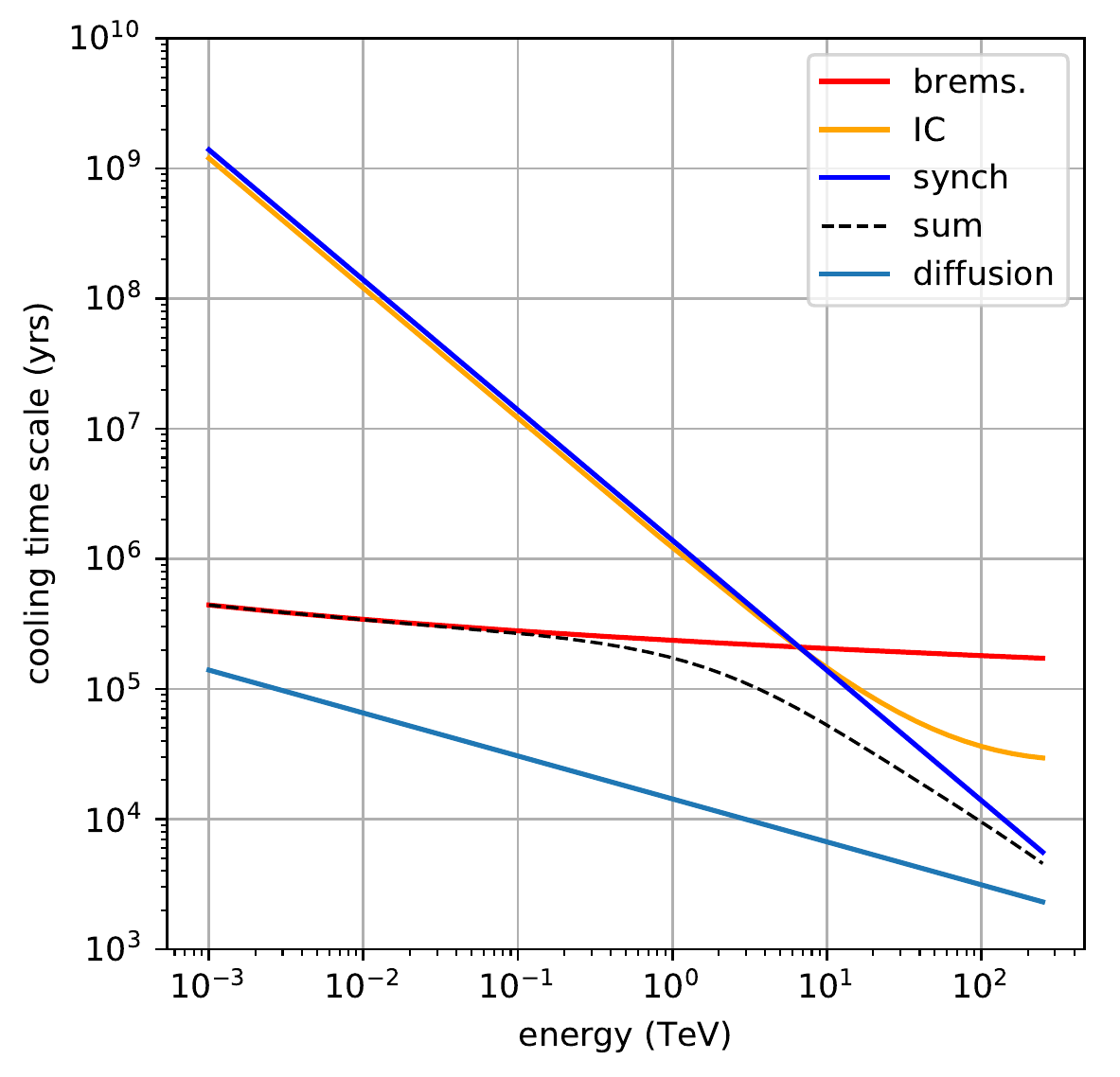}{0.5\textwidth}{(a)}
          \fig{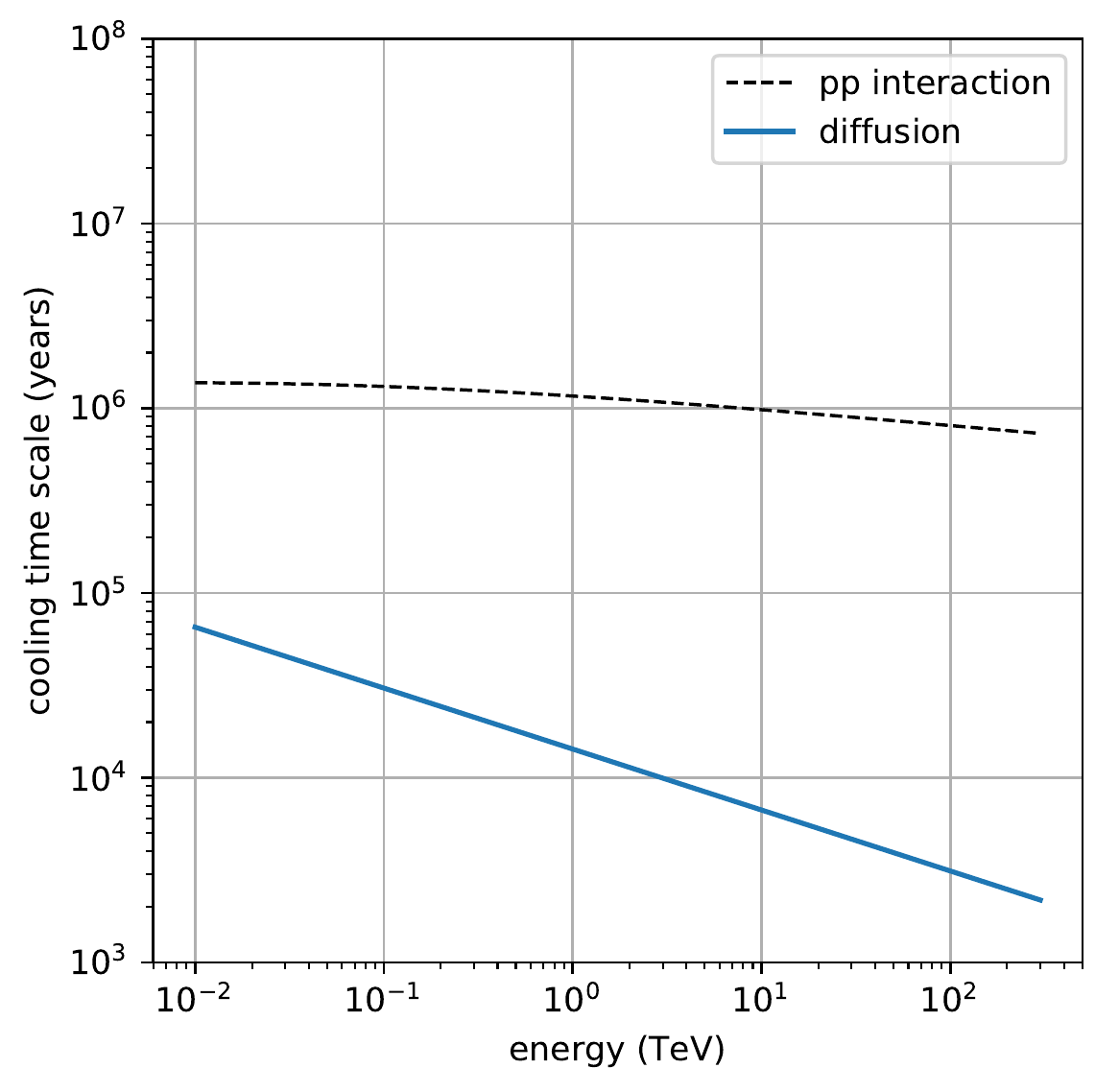}{0.5\textwidth}{(b)}
}
\caption{ Diffusion and cooling time-scales of (a) electrons due to synchrotron, inverse-Compton (IC) and bremsstruhlang emission, `sum' denotes the total cooling time-scale after including all these processes (b) protons due to proton-proton interactions for LHAASO J0341+5258
\label{fig:cooling2}}
\end{figure*}

In the above equation H denotes the height of the Galaxy and 
$D=D_0 \Big(\frac{E}{E_0}\Big)^{0.33}$ cm$^2$/sec denotes the diffusion coefficient \citep{strong2004diffuse}, with $E_0=$4 GeV. We have replaced the size of the Galaxy by the size of the molecular clouds associated with the LHAASO sources to calculate the escape time-scales during which the cosmic rays are trapped near the molecular clouds. The physical mechanism of trapping of very high energy cosmic rays near their sources is an important topic of research, for a recent review on this see \citet{marcowith2021cosmic}. 
A lower value of the constant of diffusion coefficient $D_0=10^{26}$ cm$^2$/sec compared to the value $D_0=10^{28}$ cm$^2$/sec usually adapted to explain cosmic ray propagation in our Galaxy has been used to account for the trapping of the very high energy cosmic rays in the vicinity of the LHAASO sources associated with molecular clouds. 
The masses and distances of the molecular clouds found in the vicinity of LHAASO J2108+5157 and J0341+5258 are 8469 \(M_\odot\), 3.28 kpc and 1000 \(M_\odot\), 1kpc respectively. For LHAASO J2108+5157, the average angular radius of the associated GMC is 0.236 degrees, therefore its size is about $4\times 10^{19}$ cm. The size of the molecular cloud associated with LHAASO J0341+5258 is estimated to be $1.67\times10^{19}$ cm using the masses and densities of the clouds. We have calculated the diffusion time-scales of cosmic rays from the molecular cloud regions associated with the LHAASO sources using the sizes of the clouds in Eq.(\ref{eq:difftime}).

\begin{equation}
    t_{diff}=5.07292\times10^5(\frac{E}{4 \,GeV})^{-0.33}   years
\end{equation}
for the region in the vicinity of LHAASO J2108+5157
and 
\begin{equation}
    t_{diff}=8.842\times10^4(\frac{E}{4 \,GeV})^{-0.33}  years
\end{equation}
for the region in the vicinity of LHAASO J0341+5258.
The diffusion and cooling time-scales of electrons and protons have been shown in Fig \ref{fig:cooling1} and Fig \ref{fig:cooling2}
for the regions in the vicinity of LHAASO J2108+5157 and LHAASO J0341+5258 respectively. The diffusion time scale is shorter for the cosmic rays producing gamma-rays in the case of LHAASO J0341+5258 compared to LHAASO J2108+5157. The ages of the explosions are taken to be shorter than the diffusion and cooling time-scales of the cosmic ray  electrons and protons producing the gamma-rays.

\section{Results}
LHAASO J2108+5157 \citep{2021} and LHAASO J0341+5258 \citep{cao2021discovery} are not yet found to be associated with any powerful pulsar or SNR, this motivates us to consider alternative scenarios. We have assumed there  were explosions which injected  both shock accelerated electrons and protons thousands of years ago. We have considered impulsive  injections lasting for one year with luminosities $5.5 \times 10^{40}$ erg/sec and $5 \times 10^{38}$ erg/sec in protons and electrons respectively for LHAASO J2108+5157 and luminosities $4.5 \times 10^{38}$ erg/sec and $1.5 \times 10^{38}$ erg/sec in protons and electrons respectively for LHAASO J0341+5258. We have used the spectral index of the injected cosmic ray spectrum (electrons and protons) $\frac{dN}{dE}\propto E^{-2}$. The GAMERA code has been used to calculate the time evolved cosmic ray spectrum and photon-spectrum at present day after including radiative losses of electrons, hadronic interactions and diffusion of cosmic rays. 

\begin{figure*}[b]
\gridline{\fig{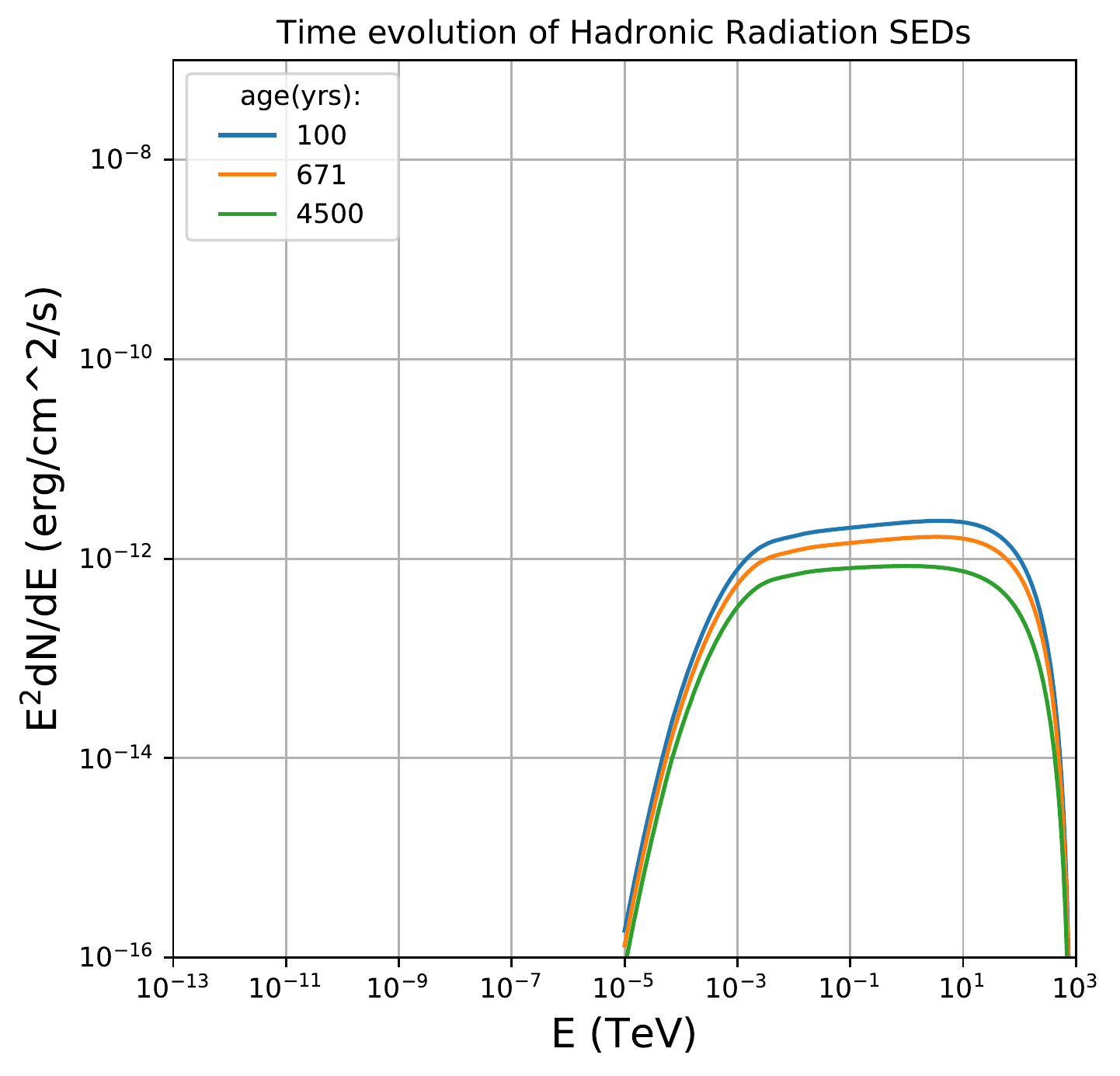}{0.5\textwidth}{(a)}
          \fig{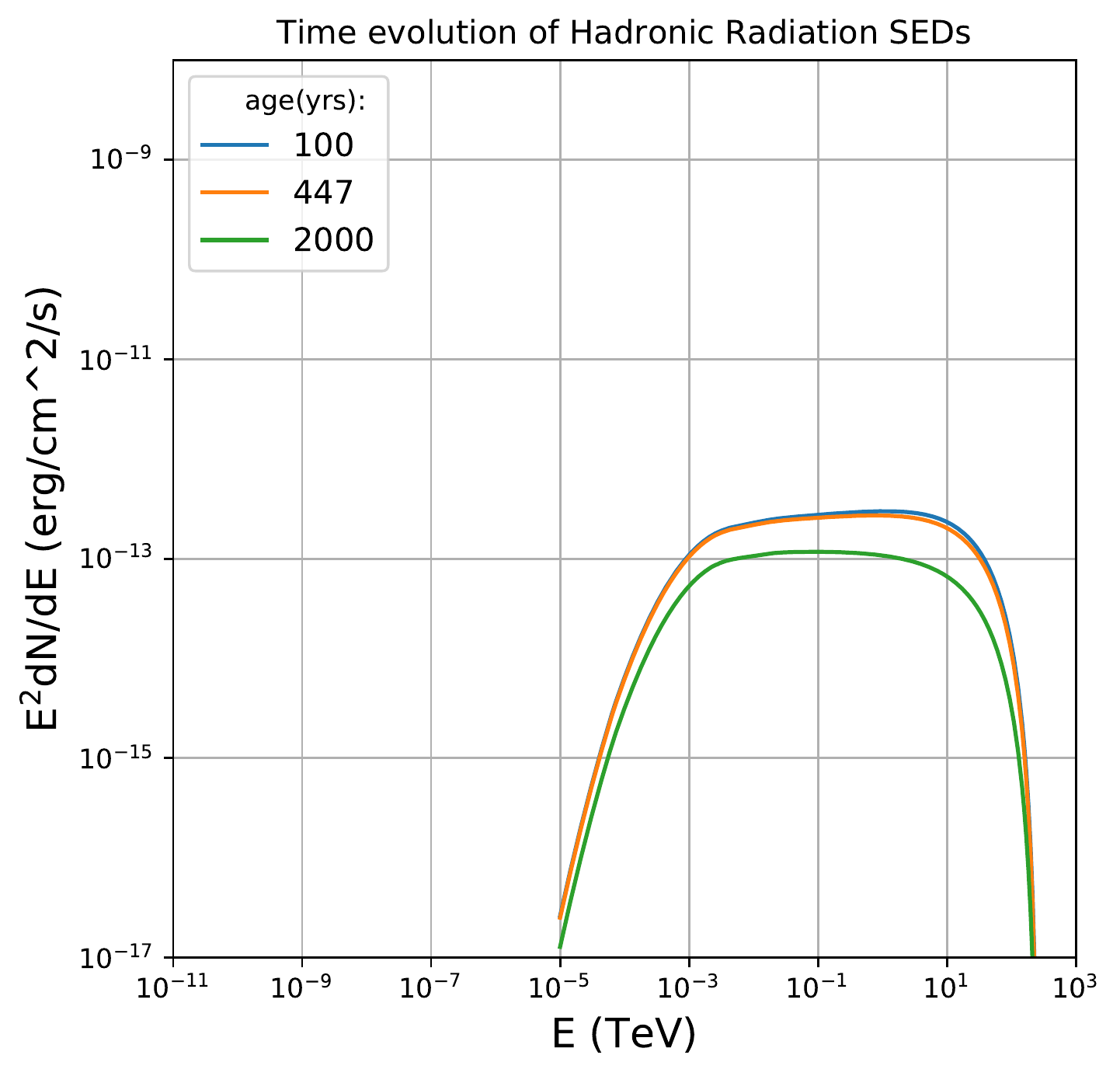}{0.5\textwidth}{(b)}
}
\caption{ Time evolution of gamma ray emission in proton-proton interactions for (a) LHAASO J2108+5157 (b) LHAASO J0341+5258
\label{fig:hadronevoltion}}
\end{figure*}
Shock accelerated protons interact with the molecular clouds present in the vicinity of the sources to emit gamma rays in proton-proton interactions. Fig \ref{fig:hadronevoltion} shows the time evolution of $\gamma$ ray flux due to proton-proton interactions for the two sources. The time intervals labeled as `age(yrs)' in the figures have been shown in equal intervals in logarithmic scale.
\begin{figure*}
\gridline{\fig{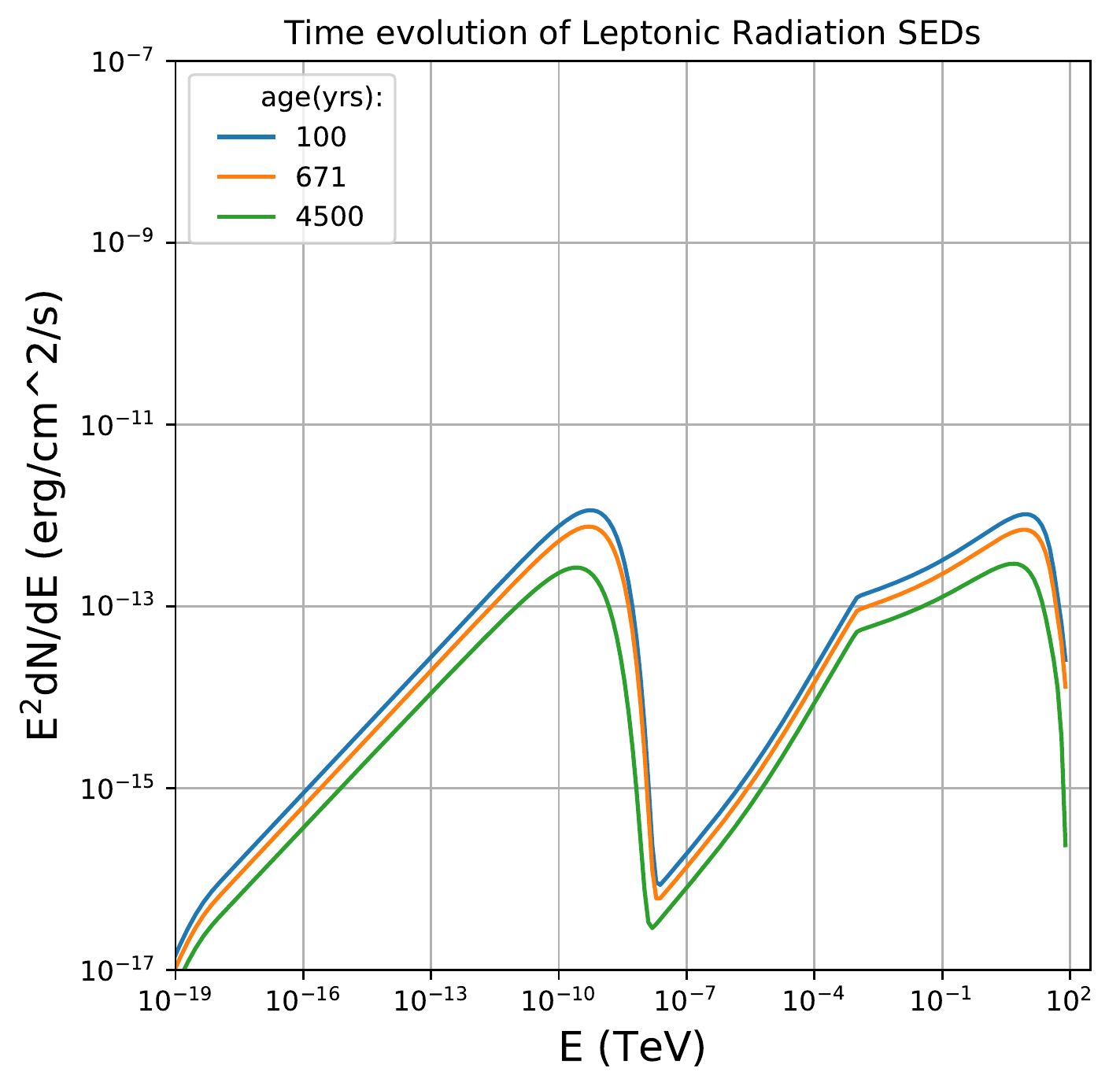}{0.5\textwidth}{(a)}
          \fig{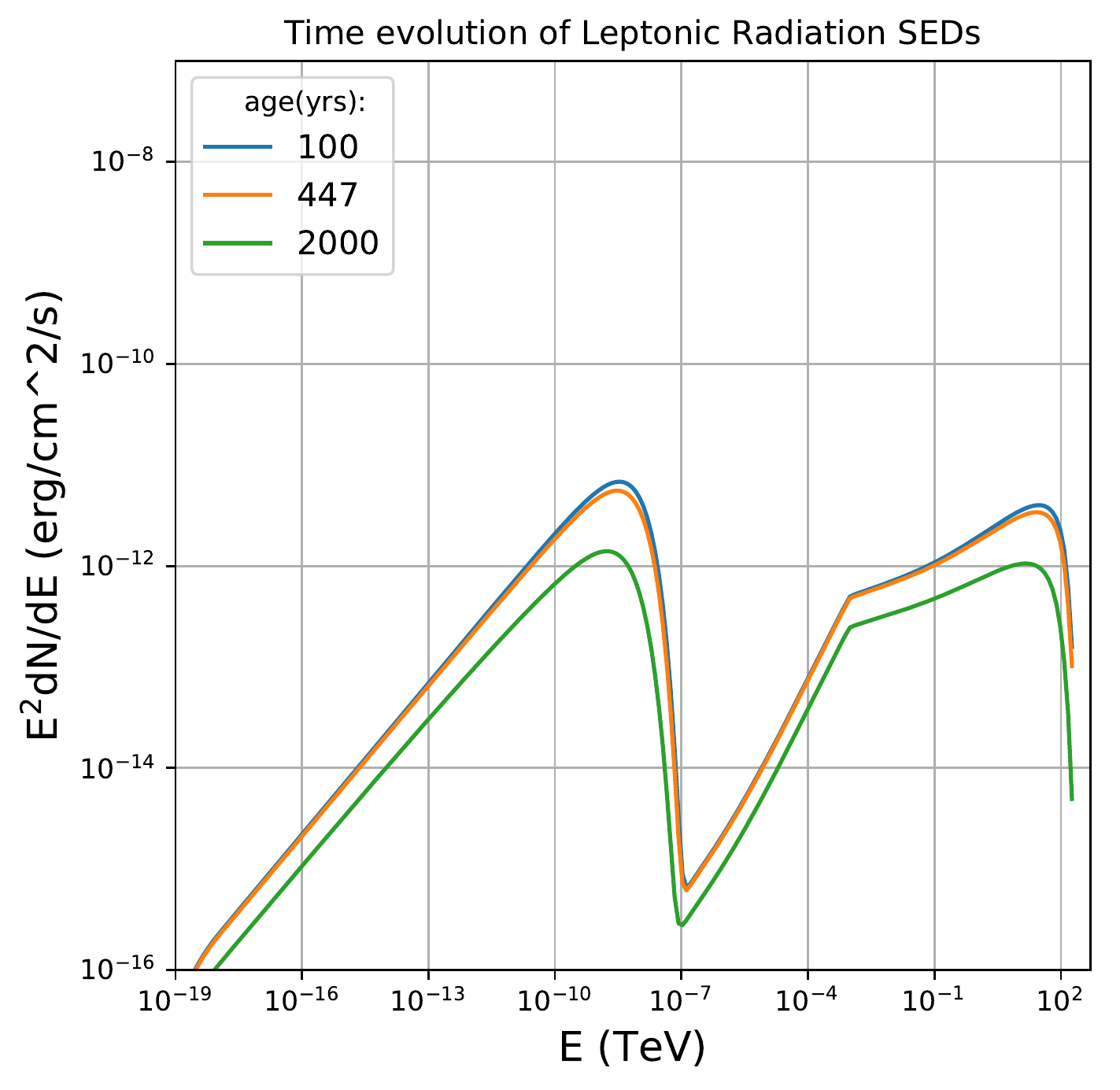}{0.5\textwidth}{(b)}
}
\caption{ Time evolution of leptonic emission including synchrotron, IC (CMB  \& ISRF), bremsstrahlung radiations for (a) LHAASO J2108+5157 (b) LHAASO J0341+5258
\label{fig:leptonevoltion}}
\end{figure*}

\begin{figure*}
\gridline{\fig{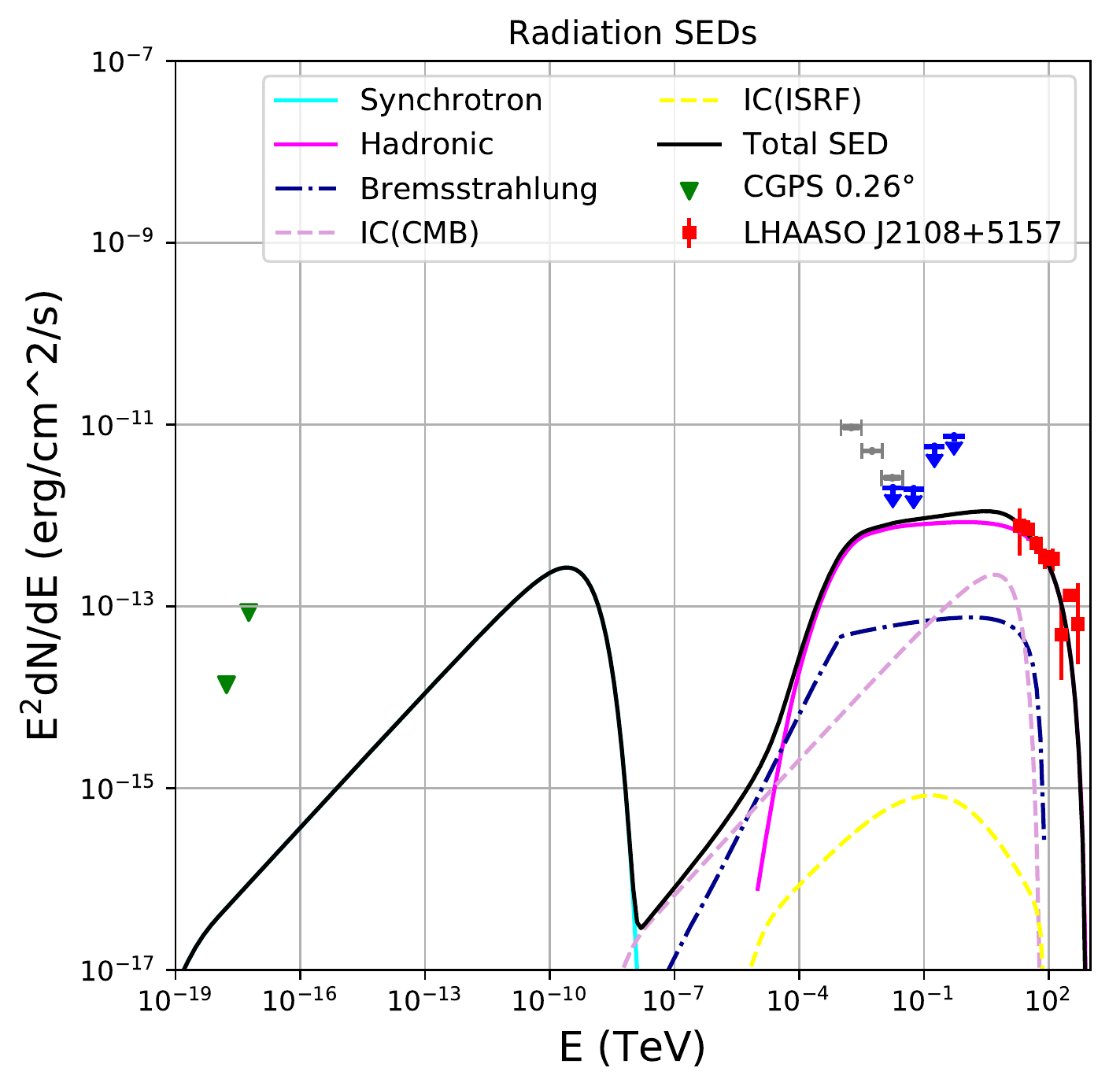}{0.5\textwidth}{(a)}
          \fig{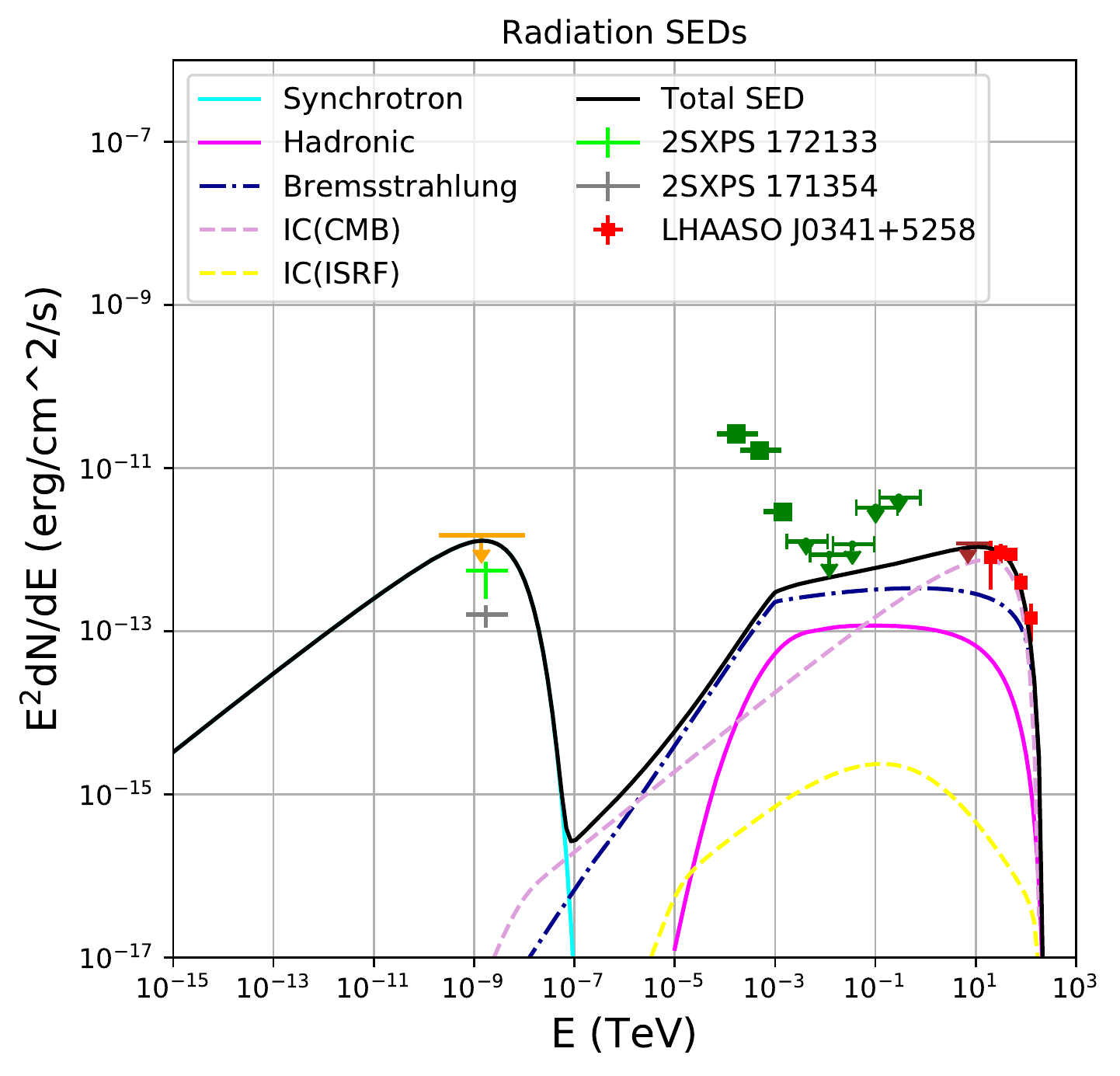}{0.5\textwidth}{(b)}
}
\caption{ 
(a)Multiwavelength SEDs LHAASO J2108+5157. Multiwavelength data points and upper limits have been shown from \citet{2021}. The blue downward arrows are the upper limits of Fermi-LAT within angular extension $\sigma= 0.26^{\circ}$ and grey points are the spectral points of Fermi-LAT within angular extension $\sigma= 0.48^{\circ}$. The green triangles are the radio flux derived by \citep{2021} within $0.26^{\circ}$ of the source, obtained from the  Canadian Galactic Plane Survey.\\(b)Multiwavelength SEDs of LHAASO J0341+5258. Multiwavelength data points and upper limits have been shown from \citet{cao2021discovery}. The orange downward arrow is the upper limit from Chandra observation, brown downward arrow is the HAWC upper limit.  The lime green and grey points are the X-ray fluxes for 2SXPS 172133 and 2SXPS 171354  respectively from the 2RXPS catalog \citep{evans2019vizier}. Green points and arrows are the Fermi-LAT spectral points and upper limits.
\label{fig:total}}
\end{figure*}

The shock accelerated electrons injected during explosions emit radiation through synchrotron, IC and bremsstrahlung. Fig \ref{fig:leptonevoltion} represents the time evolution of total leptonic contribution that is obtained by adding all the components (synchrotron, IC of background photons, bremsstrahlung) for both the sources. For IC we take into account both CMB and interstellar radiation fields (ISRFs) in the solar vicinity  \citep{popescu2017radiation} as the target photon fields.

In Fig \ref{fig:leptonevoltion} we observe that after a certain age, due to cooling of electrons the peak of the secondary photon flux slightly shifts towards lower energy.  
The radiation spectra of protons and electrons vary with the age of the explosion, and also with the total energy injected during the explosions. The values of these parameters are varied so that the total SEDs after adding the leptonic and hadronic contributions 
fits the observed spectra. In fig \ref{fig:total}, we show the total 
SEDs of the two sources along with the individual components in our lepto-hadronic model. The observed TeV spectrum of LHAASO J0341+5258 is steeper than that of LHAASO J2108+5157. The values of the parameters used in our model are listed in Table 1. We have not included absorption of UHE gamma rays by the local radiation background as the time evolved spectra are in agreement with the observed data.
The total energy at injection required to explain the observed data at present day is 1.7$\times 10^{48}$ erg and 1.9$\times 10^{46}$ erg  for LHAASO J2108+5157 and LHAASO J0341+5258, respectively.
 
\begin{deluxetable*}{ccc}
\tablenum{1}
\tablecaption{Parameters used for modelling of two sources\label{tab:parameter}}
\tablewidth{0pt}
\tablehead{
\colhead{Parameters} & \colhead{LHAASO J2108+5157} &\colhead{LHAASO J0341+5258} 
}
\decimalcolnumbers
\startdata
Energy injected in protons & $1.735\times 10^{48}$ erg & $1.4193\times 10^{46}$ erg \\ 
Energy injected in electrons &$1.577 \times 10^{46}$ erg & $4.731 \times 10^{45}$ erg \\
Maximum energy of electrons injected & $10^{2}$TeV &$250$ TeV \\
Minimum energy of electrons injected &$10^{-3}$TeV & $10^{-3}$ TeV \\
Maximum energy of protons injected &$10^{3}$TeV & $300$ TeV \\
Minimum energy of protons injected& $10^{-2}$TeV & $10^{-2}$ TeV \\
Spectral index of injected spectrum &$-2$ &$-2$\\
Magnetic field in emission region & $3 \mu $G & $3 \mu$G  \\
Number density of particles in molecular cloud & $30cm^{-3}$ & $50cm^{-3}$  \\
Distance of molecular cloud & 3.28 kpc & 1kpc  \\
Age of explosion & $4500$ years & $2000$ years \\
\enddata

\end{deluxetable*}

\section{Discussion and Conclusion}
The UHE gamma ray sources detected by HAWC and LHAASO in the last few years have drawn much attention of the gamma-ray community. Out of the dozen of UHE gamma ray sources detected by LHAASO we have selected two sources LHAASO J2108+5157 and LHAASO J0341+5258, which are not found to be associated with any powerful pulsar or SNR. We consider a scenario where  explosions in the past caused the injection of shock accelerated electrons and protons in the locations of these two sources.
We have studied the time evolution of shock accelerated electrons and protons released during explosions and their energy losses.
We have calculated the time evolved SEDs of photons in lepto-hadronic model including synchrotron, IC, bremmstrahlung and proton-proton interactions in molecular clouds. The time evolved SEDs of photons can explain the observed data at present day.
The weak supernova like explosions are expected to happen 4500 years  and 2000 years ago. The cooling time-scales of electrons and protons have been shown along with their diffusion time-scale in Fig \ref{fig:cooling1} and Fig \ref{fig:cooling2}. The explosions released energies of the order of $10^{48}$ erg and $10^{46}$ erg respectively in accelerated particles (electrons and protons).  In supernova explosions the energy emitted is typically of the order of $10^{51}$ erg, and only a few percent of it goes into non-thermal particle emission. Determining the exact nature of these explosions is beyond the scope of this paper. We conclude that supernova like explosions in the past whose reminiscence is hard to identify at present day could also be the cause of UHE gamma ray emission at present day as shown in Fig \ref{fig:total}.

\section{acknowledgment}
The authors thank Agnibha De Sarkar for helpful discussions. The authors are  thankful to the referee for invaluable comments to improve this paper. 

\software{GAMERA (\url{http://libgamera.github.io/GAMERA/docs/documentation.html})}
 
\bibliography{reference}{}
\bibliographystyle{aasjournal}

\end{document}